\documentclass[runningheads]{llncs}
\usepackage{amsmath,amsfonts}
\def\etal{\emph{et al.}}
\def\eg{\emph{e.g.}}
\usepackage{algorithmic}
\usepackage{algorithm}
\usepackage{array}
\usepackage{textcomp}
\usepackage{stfloats}
\usepackage{url}
\usepackage{verbatim}
\usepackage{graphicx}
\usepackage{cite}
\usepackage{tabularx}
\usepackage{xcolor}
\usepackage{multirow}
\usepackage{comment}
\usepackage{booktabs}
\usepackage{float}
\usepackage{hyperref}
\usepackage[T1]{fontenc}
\title{RAEDiff: Denoising Diffusion Probabilistic Models Based Reversible Adversarial Examples Self-Generation and Self-Recovery}

\author{Fan Xing\inst{1}%\orcidID{0000-1111-2222-3333} 
\and Xiaoyi Zhou\inst{1}%\orcidID{0000-1111-2222-3333} 
\and Xuefeng Fan\inst{1}
\and Zhuo Tian\inst{1}
\and Yan Zhao\inst{2}}
\authorrunning{F. Xing et al.}
\institute{The School of Cyberspace Security, Hainan University, Hainan, Haikou 570228, China \and
Science and Technology Communication Department of Hainan Provincial Public Security Department, Hainan, Haikou, China}

\begin{document}
\maketitle

\begin{abstract}
Collected and annotated datasets, which are obtained through extensive efforts, are effective for training Deep Neural Network (DNN) models. However, these datasets are susceptible to be misused by unauthorized users, resulting in infringement of Intellectual Property (IP) rights owned by the dataset creators. Reversible Adversarial Exsamples (RAE) can help to solve the issues of IP protection for datasets. RAEs are adversarial perturbed images that can be restored to the original. As a cutting-edge approach, RAE scheme can serve the purposes of preventing unauthorized users from engaging in malicious model training, as well as ensuring the legitimate usage of authorized users. %Reversible Adversarial Examples (RAE) can introduce adversarial perturbations to images, and restoring the original images with auxiliary information for eliminating perturbations.
Nevertheless, in the existing work, RAEs still rely on the embedded auxiliary information for restoration, which may compromise their adversarial abilities. %, making %the decrease of adversarial attack capability.%it unable to integrate generation and restoration. 
In this paper, a novel self-generation and self-recovery method, named as RAEDiff, is introduced for generating RAEs based on a Denoising Diffusion Probabilistic Models (DDPM). It diffuses datasets into a Biased Gaussian Distribution (BGD) and utilizes the prior knowledge of the DDPM for generating and recovering RAEs.%Experiments are conducted to investigate the impact of RAEs on Artificial Intelligence Generated Content (AIGC) models for the first time. 
The experimental results demonstrate that RAEDiff effectively self-generates adversarial perturbations for DNN models, including Artificial Intelligence Generated Content (AIGC) models, while also exhibiting significant self-recovery capabilities.%Experiments are conducted on CIFAR-10 and CelebA datasets and, for the first time, investigate the impact of RAEs on Artificial Intelligence Generated Content (AIGC) models. The results demonstrate that RAEDiff effectively self-generates adversarial effects for DNN models, including AIGC models, while also exhibiting significant self-recovery capabilities.
\end{abstract}

\keywords{Diffusion model \and Reversible adversarial examples \and Intellectual property protection}
    
\section{Introduction}
\label{sec:intro}

\begin{figure}[ht]
  \centering
   \includegraphics[width=0.5\linewidth]{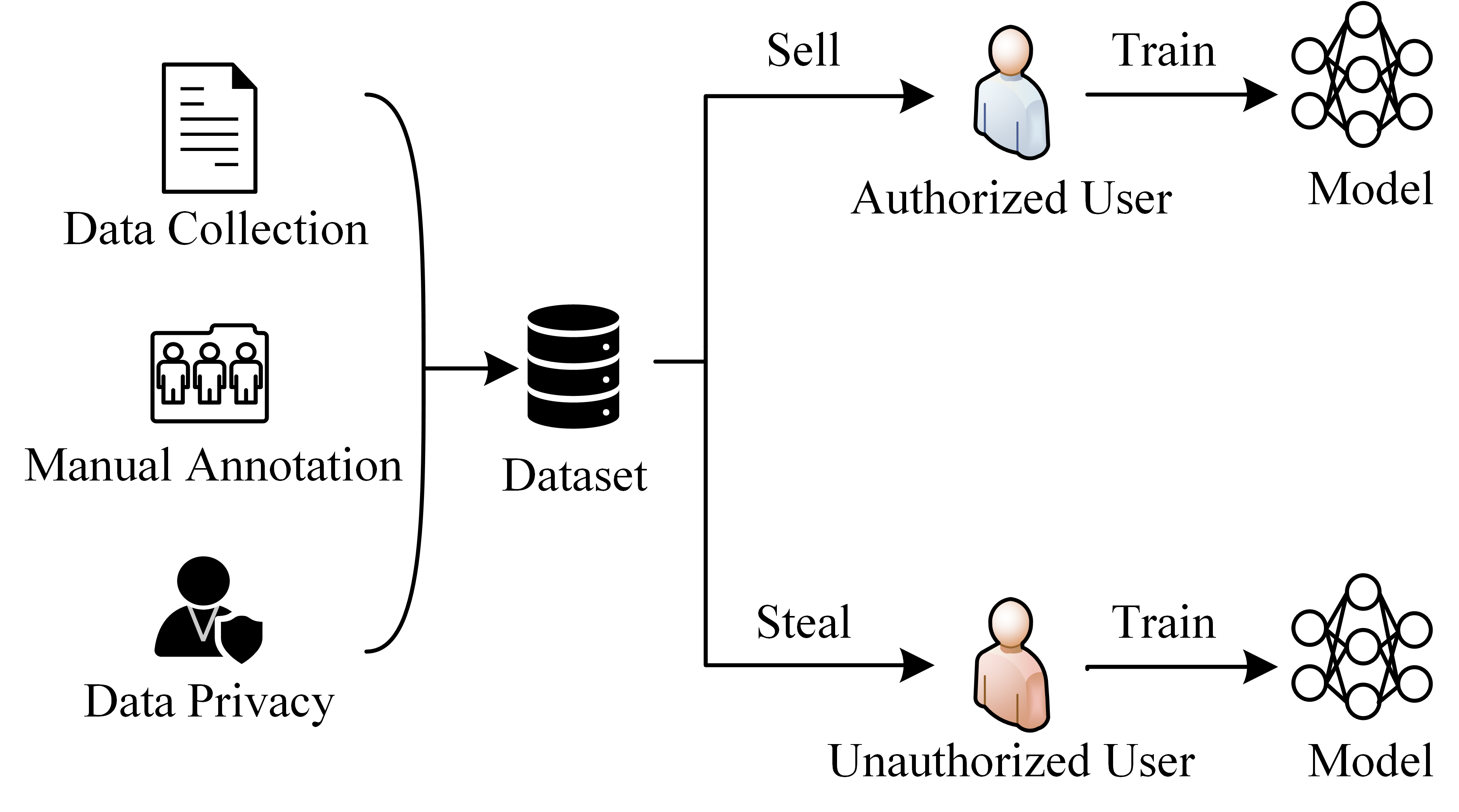}
   \caption{Example of dataset infringement.}
   \label{fig:1}
\end{figure}

The performance of Deep Neural Networks (DNN) \cite{1,2,3,4,5}is heavily dependent on the quantity and quality of the training data. Artificial Intelligence Generated Content (AIGC), one of the most outstanding achievements in DNN, enabling DNN to generate content based on specific requirements and requirements, and has extensive practical applications in areas such as image synthesis and language translation.%\cite{1,2,3,4,5}1,5
However, as shown in Figure \ref{fig:1}, malicious users may steal datasets for training DNN models without authorization from the dataset owners, potentially raising significant Intellectual Property (IP) concerns \cite{6,7,9,32}.%\cite{6,7,9,32}6,9
As a result, a critical question needs to be considered: Can users' images be maliciously used for DNN model training without authorization?
In response to these concerns, Liu \etal \cite{36} introduced the concept of Reversible Adversarial Examples (RAE). By generating adversarial examples tailored explicitly to the images, it ensures that the generation process of the adversarial examples is reversible, allowing authorized DNN models to restore the images to their original state. Applying an RAE scheme to IP protection of dataset is considered an extremely effective means for preventing malicious training on the dataset \cite{21}.%（这句话应该要有引用。对于重要结论的，不是自己得出的结论，都务必有引用才有可靠性,已加上）
Moreover, in view of practical application scenarios, to %facilitate individual users to %
train high-performance cloud-based models on Machine Learning as a Service (MLaaS) \cite{49}, RAEs should be compatible with Data Management Cloud Platforms (DMCP) and be able to recover and clean the datasets \cite{21}. %Subsequently, the models can be trained on Machine Learning as a Service (MLaaS) \cite{49}.
The existing RAE methods pose certain challenges as follows \cite{36,19,20,21,22}: 
\begin{itemize}
    \item Challenge 1: Reversible Data Hiding (RDH) \cite{45,13,14} embeds the auxiliary information into RAEs for the restoration while preserving the images' quality, the RAE methods are predominantly leveraging RDH to achieve reversibility. However, the embedding of auxiliary information leads to the decrease of attack capability, and might make it difficult to restore.%\cite{45,10,11,12,13,14}
    % \item Most of the existing RAE research has predominantly focused on the expansion of a limited number of images rather than addressing entire datasets.
    \item Challenge 2: Taking into account real-world application contexts, the current work \cite{18,19,20,21,22,36} lacks the capability to differentiate RAEs based on training performance due to the diverse requirements of users in DMCP. 
    
\end{itemize}

Denoising Diffusion Probabilistic Models (DDPM)\cite{40} is one of the widely used AIGC models for image generation. A well-trained DDPM iteratively denoises images from a randomly sampled noise through the utilization of prior knowledge to enhance the semantic representation of images \cite{52repaint,53ddpm}.
Inspired by the backdoor attacks on DDPM \cite{28,29}, biasing the training process of DDPM to alternative Gaussian distributions is unlikely to significantly impact the model's original performance.%(这里用motivate值得商榷） 

For the purpose of overcoming Challenge 1, we designed an Adversarial Generative Process (AGP) to mislead the standard generative process of enhancing semantic representation in DDPM to generate adversarial perturbations. Since all perturbations are generated by such DDPM, it is possible to recover the perturbations without any auxiliary information through a standard generative process using a well-trained DDPM. Finally, the DDPM is biased to a Gaussian distribution controlled by the backdoor trigger, enabling the trigger to serve as a distinctive watermark within the noise introduced during AGP.
% From the above work, to generate adversarial perturbations, we designed an Adversarial Generative Process (AGP) to mislead the standard generative process of enhancing semantic representation in DDPM. 

To tackle Challenge 2, \textbf{Dataset Training Performance Permission Management (DTPPM)} is introduced to provide datasets with varying training performances tailored to different requirements.%(上面没提到limitation，在这里突然出现不合适，in the second challenge）(为了方便引用，将上面的challenges分别用专用词替代，现在出现了一个问题，Challenge 1在哪里体现了？根据论文逻辑，第24段应该要讲如何应对挑战1）
An RAE-based method, named \textbf{RAEDiff}, is proposed in this study. Unlike the existing RAE methods, it enables self-generation and self-recovery of RAEs merely by a trained DDPM without any auxiliary information embedding which might impact the adversarial capability.

Our contributions are as follows:
\begin{itemize}
\item \textbf{RAEDiff} is introduced for the first implementation of generating RAEs through self-generation and self-recovery for IP protection of DNN image datasets.
In contrast to existing RAE methods that rely on %adversarial attack for generation and 
auxiliary information for restoration, \textbf{RAEDiff} attempts to unify the RAE generation and restoration by utilizing the devised AGP and leveraging the prior knowledge of the DDPM.

\item The training performance requirements and the commercial value of datasets are taken into account when considering the real-world application of RAEs on DMCP. Therefore, DTPPM is proposed to provide varying training performance across multiple domains and application scenarios. 

\item Experiments demonstrate the feasibility of \textbf{RAEDiff} to generate RAEs for protecting DNN image datasets, which ensures the availability of the dataset for authorized users and effectively prevents unauthorized users from illegal usage.
\end{itemize}

The rest of this paper is organized as follows.
The related work are reviewed in Section \ref{sec:formatting}. 
% The background and problem definition is described in Section \hyperref[sec:prodefin]{3}.
\textbf{RAEDiff} is illustrated in Section \ref{sec:method}.
The evaluation of the proposed method is displayed in Section \ref{sec:expri}.
This paper is concluded in Section \ref{sec:conclu}.

%-------------------------------------------------------------------------

\section{Related work}
\label{sec:formatting}
% \subsection{Reversible Data Hidding}
% RDH plays a crucial role in protecting the intellectual property (IP) of image datasets while maintaining hidden data reversibility.
% Currently, research on RDH has matured and shown outstanding results \cite{10,11,12,13,14}.
% \begin{comment}
% Jia \etal \cite{10} proposed an RDH scheme based on image texture to reduce invalid pixel shifts in histogram shifting.
% Mao \etal \cite{11} proposed an efficient color image RDH scheme based on pixel value ordering (PVO), leveraging channel correlation to enhance embedding performance.
% Zhao \etal \cite{12} presented a histogram shifting technique that allows overlapping between multiple pairs of maximum and minimum points.    
% \end{comment}
% Fu \etal \cite{13} proposed a denoiser for preprocessing watermarks, enhancing robustness against geometric and conventional attacks, reducing the amount of extracted auxiliary information, and designing a more stable auxiliary information extraction process.
% Zhou \etal \cite{14} introduce a new approach in JPEG reversible data hiding using pairwise modification which focuses on optimizing 2D mappings and ordering strategies by carefully initializing pairing rules.
% However, the training values of image datasets for DNN models have not been fully considered by RDH.
%-------------------------------------------------------------------------------------------------------------------
% \subsection{Reversible Adversarial Example}
\subsection{Backdoor attack on DDPM}
Research on backdoor attacks against DDPM is still in its infancy \cite{28,29}, but it yields excellent results. The original and the backdoor attack tasks are trained under two different Gaussian distributions. A trigger is employed to shift the original task's distribution towards that of the backdoor attack task, resulting in the model generating outputs influenced by the backdoor attacks.
\subsection{Reversible Adversarial Example}
Relevant experts have explored numerous different solutions to generate RAEs \cite{18,19,20,21,22,36} for IP protection of datasets, but they all require auxiliary information for restoration. %We have compared these existing works and our proposed method shown in Table \hyperref[tab:workscompare]{1}.
Mao \etal \cite{18} have discovered that images contain inherent structures that allow adversarial attacks to be reversed.
In pursuit of reversibility in adversarial attacks, Liu \etal \cite{36} first introduced the RAE method based on Reversible Data Hidding (RDH) and demonstrated its feasibility.
When exploring alternative methods for generating RAEs, Yin \etal \cite{19} first proposed a reversible adversarial attack based on Reversible Image Transformation (RIT) \cite{51rit} and pointed out that if adversarial examples possess both attack capability and reversibility, they undoubtedly hold significant practical value.
To address the passive state in dataset IP verification, Xue \etal \cite{21} first proposed an active methodology for generating RAEs based on RIT. Notably, this is the sole work focusing on the generation of RAEs within the dataset.
To achieve better attack capability and visual quality of RAEs, Yin \etal \cite{20} proposed a method based on the independence of luminance and chrominance channels in the YUV color space.
With the aim of more comprehensively examining the application of RAE in real scenarios, Xiong \etal \cite{22} proposed a Perturbation Generative Network (PGN) that could generate RAEs in a black-box scenario.

%Zhang \etal \cite{22} proposed a recoverable Generative Adversarial Network (GAN) for generating RAE. They utilized a dimensionality reducer to optimize the distribution of adversarial perturbations.

% \begin{table}[h]
%     \centering
%     \begin{tabular}{c|c}
%          \toprule
%          Works &  Auxiliary information embedding\\
%          \midrule
%          \cite{36} &  RDH required \\
         
%          \cite{19} &  RDH required\\
         
%          \cite{21} &  RDH required \\
         
%          \cite{20} &  RDH required \\
         
%          \cite{22} &  RDH required \\
         
%          Ours & Not require\\
%          \bottomrule
%     \end{tabular}
%     \caption{The existing works and our proposed method are compared in terms of whether they require RDH to embed auxiliary information for RAE restoration}
%     \label{tab:workscompare}
% \end{table}

In summary, there have been remarkable research achievements on RAE schemes. However, it is proved that the embedding of auxiliary information significantly impacts adversarial performances \cite{19,22}, while existing work has not taken into account the training effectiveness of AIGC models. Therefore, aimed to generate and restore RAEs without any transformation and auxiliary information, an entirely novel approach is explored in this paper. Experiments are conducted to investigate the impact of RAEs on AIGC models.
%-------------------------------------------------------------------------------------------------------------------

 % As a result, this further emphasizes the need for RAE-based dataset IP protection to focus on preventing effective training of DNN models.  Therefore, we propose a new diffusion model based method for protecting datasets' IP and to address the above issues 
%-------------------------------------------------------------------------

%-------------------------------------------------------------------------
%\input{sec/3_finalcopy}
% \input{sec/4_method}
\section{Experiments}
\label{sec:expri}
%----------------------------------------------------------------------------------
\subsection{Experiment setup}
\subsubsection{Datasets, models, and implementation details}
\noindent We utilize two commonly used datasets: ten-class dataset CIFAR-10 (32 × 32) \cite{38} and facial dataset CelebA (64 × 64) \cite{39}. We followed \cite{47,48} and combined the three most balanced attributes of CelebA, namely Heavy Makeup, Mouth Slightly Open, and Smiling, into eight distinct classes. We adopt the diffusion models DDPM \cite{40} and ResNet-18\cite{50res18}, and follow their structures and training details. Our experiments are conducted on the basis of \cite{28}. 50k samples on CIFAR-10 and 10k on CelebA are sampled respectively for the evaluation of performance. In order to maximize the quality of the recovered images, we trained DDPM from scratch for 500K iterations until the generated image quality approached the baseline \cite{28}. For the provided triggers $\delta$ by authorized users, we selected Hello Kitty image, with scale factor $\gamma$ set to 0.6.\\
\subsubsection{Evaluation metrics}
Four widely-used metrics for image quality are selected for evaluation, i.e., Frechet Inception Distance(FID) \cite{41}, Learned Perceptual Image Patch Similarity(LPIPS) \cite{44}, Structure Similarity Index Measure(SSIM) \cite{55ssim} and Peak Signal-to-Noise Ratio(PSNR). A lower FID indicates a better reality of the image for humans, and LPIPS is employed to assess the differences between images from human perception.
%----------------------------------------------------------------------------------
\subsection{Ablation Studies}
\begin{figure}[htbp]
    \centering
    \includegraphics[width=0.7\linewidth]{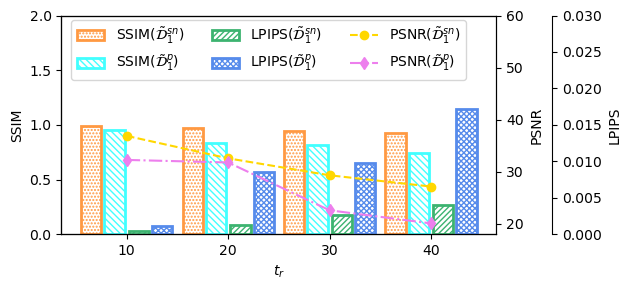}
    \caption{Visual evaluation of $\Tilde{\mathcal{D}}_1^{s n}$ and ${\mathcal{D}}_1^{p}$ generated from different $t_r$ when $\eta$ is set to 1.4.% When $t_r=30$, it is evident that the PSNR values of $\Tilde{\mathcal{D}}_1^{s n}$ and ${\mathcal{D}}_1^{p}$ decrease sharply, and LPIPS values significantly increase. %Therefore, when $t_r$ takes on excessively large values, it leads to a deterioration in the restoration performance from ${\mathcal{D}}_1^{p}$ to $\Tilde{\mathcal{D}}_1^{s n}$. 
    %When $t_r = 10$, $\Tilde{\mathcal{D}}_1^{p}$ contains very little noise, resulting in low adversarial capability.
    }
    \label{fig:trN}
\end{figure}

The visual performance impact on RAEs of hyperparameters $t_r$ and $\eta$ are investigated, applied within \textbf{RAEDiff}, on the quality of protected dataset generation and restoration.
As shown in Figure \ref{fig:trN}, when $t_r$ is too large (\eg, 40), it is challenging to recover ${\mathcal{D}}_1^{p}$. On the other hand, when $t_r$ is too small (\eg, 10), despite achieving better recovery quality,  ${\mathcal{D}}_1^{p}$ might lack adversarial perturbations.

\begin{figure}[htbp]
    \centering
    \includegraphics[width=0.7\linewidth]{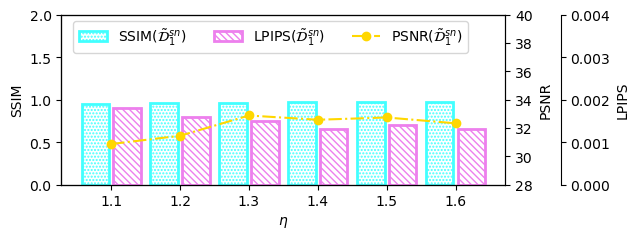}
    \caption{Visual evaluation of $\Tilde{\mathcal{D}}_1^{s n}$ generated from different $\eta$ when $t_r$ is set to 20.}
    \label{fig:etaN}
\end{figure}

When $t_r$ is held constant, the generated ${\mathcal{D}}_1^{p}$ remains unchanged.
Figure \ref{fig:etaN} illustrates that as $\eta$ increases, the time steps for restoration also increases. If $\eta$ is very large (\eg, 1.6), it will be over-denoising, which is most prominently observed in the reduction of PSNR. Conversely, it could result in an inability to achieve effective restoration.
% \begin{figure*}[htbp]
%     \centering
%     \includegraphics[width=1\linewidth]{trheeta.png}
%     \caption{Performance of $\Tilde{\mathcal{D}}_1^{s n}$ and ${\mathcal{D}}_1^{p}$ generated from different $t_r$ along with ${\mathcal{D}}_1^{p}$ generated from different $\eta$ .}
%     \label{fig:trheeta}
% \end{figure*}\\
\\
%----------------------------------------------------------------------------------
\subsubsection{The selection of $t^{s n}_m$, $t_r$ and $\eta$}
By default, the maximum time step $\mathrm{T}$ of 
 DDPM is set to 1000. Therefore, we divided the interval of $\beta_t$, ranging from a minimum of 0.0001 to a maximum of 0.02, into 1000 segments using the linear approach. This allowed us to calculate the value of $\alpha_t$. After culculating the $\bar{\alpha}_t=\prod_{i=1}^t \alpha_i$, we obtain $\bar{\alpha}_1=0.9990, \bar{\alpha}_2=0.9980, \bar{\alpha}_1=0.9960, \ldots, \bar{\alpha}_T=4.0358 \mathrm{E}{-}05$. According to Equation \eqref{1}, it can be seen that the larger $\bar{\alpha}_t$ is, the more noise the diffusion process introduces, and the smaller the proportion of the original image becomes. For this reason, to introduce slight noise to the dataset for grading training performance while ensuring the integrity of the data, we choose $t_m^{s n}=1,2$ and 3 corresponding to the permission levels 1,2, and 3. Moreover, in order to avoid introducing additional noise and thereby increasing the difficulty of recovering the image, $t_r$ is set to 20 and $\eta$ to 1.4.\\
%----------------------------------------------------------------------------------
\subsection{Experimental Results}
In this section, the experimental results of our proposed method are presented. Additionally, we conducted an ablation study on the adverse time step $t_r$ and reverse factor $\eta$.\\
\subsubsection{Visualization results}
\begin{figure*}[htbp]
     \centering
     \begin{tabular}{*{12}{p{1cm}}}
     \toprule
          \multicolumn{6}{c}{\textbf{CIFAR-10}}&  \multicolumn{6}{c}{\textbf{CelebA}}\\
          \midrule
          % $\mathcal{D}$& $\mathcal{D}_1^{s n}$& $\mathcal{D}_2^{s n}$& $\mathcal{D}_3^{s n}$& $\Tilde{\mathcal{D}}_1^{s n}$ & $\mathcal{D}_1^p       $ 
          % & $\mathcal{D}$& $\mathcal{D}_1^{s n}$& $\mathcal{D}_2^{s n}$& $\mathcal{D}_3^{s n}$& $\Tilde{\mathcal{D}}_1^{s n}$ & $\mathcal{D}_1^p$ \\
          % \midrule
         \multicolumn{6}{c}{\includegraphics[width=0.5\linewidth]{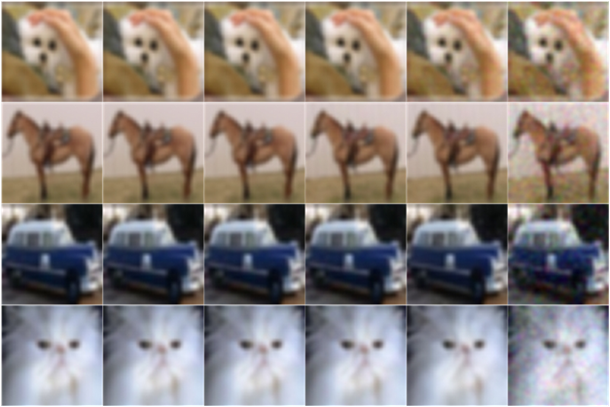}}
         & \multicolumn{6}{c}{\includegraphics[width=0.5\linewidth]{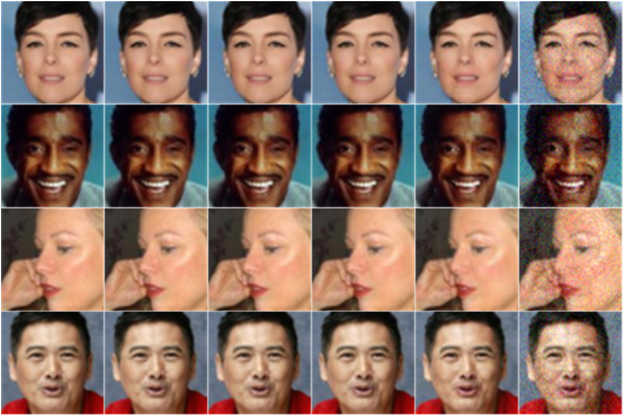}}
     \end{tabular}

     \caption{Visualization of images from $\mathcal{D}$, $\mathcal{D}_1^{s n}$, $\mathcal{D}_2^{s n}$, $\mathcal{D}_3^{s n}$, $\Tilde{\mathcal{D}}_1^{s n}$ and $\mathcal{D}_1^p$ respectively on CIFAR-10 and CelebA, arranged from left to right.}
     \label{fig:cifar10duibi}
\end{figure*}

\begin{figure*}[htbp]
    \centering
    \includegraphics[width=1\linewidth]{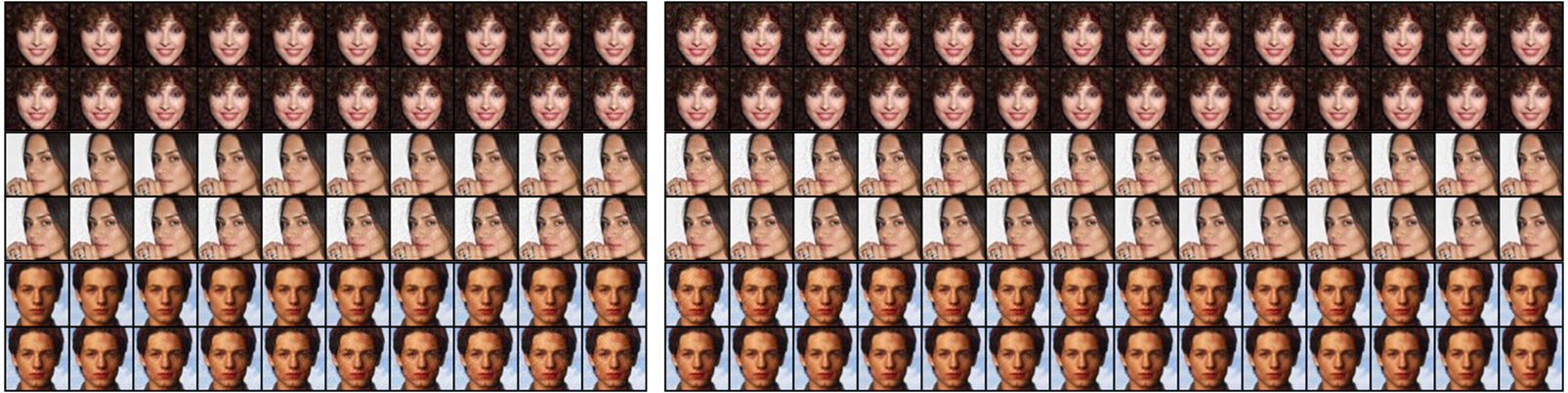}
    \caption{Visualization of the RAE generation and restoration process. The left side of each image represents the RAE generation, while the right side depicts RAE restoration.}
    \label{fig:generationrae}
\end{figure*}

The visualization of protected datasets generation and restoration are presented in this section. As depicted in Figure \ref{fig:cifar10duibi}, the protected datasets manage to maintain the quality of the original images to a significant extent while introducing adversarial noise. Additionally, the disparity between the restored images and the original images is quite small. The process of generation and restoration are shown in Figure \ref{fig:generationrae}. %and \hyperref[fig:restroationrae]{8}.
%and \hyperref[fig:celebaduibi]{6}
% \begin{figure*}[htbp]
%     \centering
%     % \includegraphics[width=0.48\linewidth]{generationrae.png}
%     % \includegraphics[width=0.48\linewidth]{restorationrae.png}
%     \includegraphics[width=1\linewidth]{new_generaionrae.png}
%     \caption{Visualization of the RAE generation and restoration process. The left side of each image represents the RAE generation, while the right side depicts RAE restoration.}
%     \label{fig:generationrae}
% \end{figure*}
%2023/8/25 实验部分
%----------------------------------------------------------------------------------
% \begin{table}[]
%     \centering
%     \renewcommand{\arraystretch}{1.2}
%     \begin{tabular}{c|c|c}
%          \toprule
%          & {\cellcolor{mygray}\textbf{CIFAR-10}} & {{\cellcolor{mygray}\textbf{CelebA}}} \\
%          \midrule
%          Clean $\mathcal{D}_1^{s n}$& 0.33 & 0.69 \\
%          \midrule
%          Protected $\mathcal{D}_1^p$& 85.48 & 62.63\\
%          \midrule
%          Recovered $\Tilde{\mathcal{D}}_1^{s n}$& 1.00 & 3.92 \\
%          \bottomrule
%     \end{tabular}
%     \caption{The FID of $\mathcal{D}_1^{s n}$, $\mathcal{D}_1^p$, $\mathcal{D}_1^{s n}$ on CIFAR-10 and CelebA.}
%     \label{tab:fidlevel1}
% \end{table}\\
%----------------------------------------------------------------------------------
\subsubsection{Accuracy}
The dataset with permission level 1 $\mathcal{D}_1^{s n}$ and its protected dataset $\mathcal{D}_1^p$ along with its recovered dataset $\Tilde{\mathcal{D}}_1^{s n}$ are chosen for evaluation. We followed the approach of Xue \etal \cite{21}, where we trained ResNet-18 from scratch.
In Table \ref{tab:accxue}, we presented a comparison between the training and testing accuracies of ResNet-18 trained directly on the original dataset. It can be observed that the model trained on $\mathcal{D}_1^p$ experiences a sharp decline in performance, dropping from a testing accuracy of 94.67\% to 42.75\% on Cifar-10. However, $\Tilde{\mathcal{D}}_1^{s n}$ can restore the original performance to a test accuracy of 94.68\%. The impact on the performance of $\mathcal{D}_1^p$ is more pronounced on CelebA, declining from a test accuracy of 81.39\% to 29.40\%. However, the original test accuracy can still be recovered on $\Tilde{\mathcal{D}}_1^{s n}$. These demonstrate that the RAEs generated by \textbf{RAEDiff} can mislead the model training and be self-restored effectively.
\begin{table*}[htbp]
    \centering
    \renewcommand{\arraystretch}{1}
    \caption{The training accuracy and test accuracy of the ResNet-18 model on clean, protected and recovered datasets trained on CIFAR-10 and CelebA.}
    %0.47
    \label{tab:accxue}
    \resizebox{0.9\textwidth}{!}{
    \begin{tabular}{c|c|cccc}
         \toprule
         \multicolumn{2}{c|}{} & \multicolumn{2}{c}{\textbf{CIFAR-10}} & \multicolumn{2}{c}{\textbf{CelebA}} \\
         \midrule
         \multicolumn{2}{c|}{\textbf{Method}} 
         &{\textbf{Training Acc}} & {\textbf{Testing Acc}}& {\textbf{Training Acc}} & {\textbf{Testing Acc}}\\
         \midrule
          & Clean & 96.44\% & 86.21\% & - & - \\
        Xue \etal \cite{21} & Protected & 100.00\% & 38.23\% & - & -\\
         & Recovered & 95.21\% & 85.86\%& - & - \\
         \midrule
          & Clean $\mathcal{D}_1^{s n}$& 97.43\% & 94.67\%              &82.86\% &81.39\%  \\
        Ours & Protected $\mathcal{D}_1^p$     & 99.08\% & 42.75\%              &76.05\% &29.40\%   \\
         & Recovered $\Tilde{\mathcal{D}}_1^{s n}$ & 99.12\% & 94.68\%      &81.78\% &80.76\%    \\
         \bottomrule
    \end{tabular}}
    
\end{table*}
%----------------------------------------------------------------------------------
\subsubsection{Visual evaluation metrics}
We have evaluated proposed datasets $\mathcal{D}_m^{s n}, \mathcal{D}_m^p, \tilde{\mathcal{D}}_m^{s n}$ by measuring their average SSIM, PSNR, and LPIPS values in Table \ref{tab:ssimpsnrlpips}. As mentioned in Section \ref{generation RAE}, the $\mathcal{D}_m^{s n}$ dataset is introduced slight, randomly sampled noise into the original images at different time steps diffusion process, with the purpose of distinguishing varying training performances.
For the $\mathcal{D}_1^{s n}$, $\mathcal{D}_2^{s n}$, and $\mathcal{D}_3^{s n}$ on CIFAR-10, their SSIM values are 0.997, 0.994, and 0.992, respectively. This demonstrates that the images processed through diffusion processes at different time steps can be clearly distinguished from each other.
On CIFAR-10, the average SSIM value of $\mathcal{D}_1^{p}$ is 0.832. It suggests that the RAEs generated by \textbf{RAEDiff} has a relatively minor impact on the original quality. In contrast, the average SSIM value of $\Tilde{\mathcal{D}}_1^{s n}$ indicates a significant restoration of the original image's structure. Furthermore, the LPIPS values for both $\mathcal{D}_1^{p}$ and $\Tilde{\mathcal{D}}_1^{s n}$ are within an extremely small numerical range (0.0086 and 0.0013, respectively). Consequently, this signifies that both $\mathcal{D}_1^{p}$ and $\Tilde{\mathcal{D}}_1^{s n}$ possess differences from the original images that are hardly perceptible to the human.
For $\mathcal{D}_1^{p}$, the changes in SSIM, PSNR, and LPIPS in CIFAR-10 are smaller compared to CelebA. We attribute this to the fact that CelebA images are larger and more complex than CIFAR-10. Nonetheless, \textbf{RAEDiff} has demonstrated its strong reversibility, as evidenced by the high SSIM on $\Tilde{\mathcal{D}}_1^{s n}$.
\begin{table*}[htbp]
    \centering
    \caption{The average SSIM, PSNR and LPIPS of $\mathcal{D}_1^{s n}$, $\mathcal{D}_2^{s n}$, $\mathcal{D}_3^{s n}$, $\mathcal{D}_1^{p}$, $\Tilde{\mathcal{D}}_1^{s n}$ and $\Tilde{\mathcal{D}}_1^{atk}$ on CIFAR-10 and CelebA.}
    \label{tab:ssimpsnrlpips}%0.93 48
    \resizebox{0.8\textwidth}{!}{
    \begin{tabular}{c|c|cccccc}
         \toprule
         % \multicolumn{2}{c|}{} 
        & \textbf{Metrics}& \textbf{$\mathcal{D}_1^{s n}$} & {\textbf{$\mathcal{D}_2^{s n}$}} & \textbf{$\mathcal{D}_3^{s n}$} & \textbf{$\mathcal{D}_1^{p}$} & \textbf{$\Tilde{\mathcal{D}}_1^{s n}$} & {{\textbf{$\Tilde{\mathcal{D}}_1^{a t k}$}}}\\
         \midrule
         &SSIM&0.997&0.994&0.992&0.832&0.969 &0.787\\
CIFAR-10&PSNR&48.90&45.73&43.68&31.79&32.58&27.96\\
                &LPIPS&1.288E-05&4.515E-05&7.067E-05&0.0086&0.0013&0.0121\\
        \midrule
        	&SSIM&0.990&0.983&0.977&0.689&0.945&-\\
         CelebA       &PSNR&48.73&45.54&43.48&24.20&34.70&-\\
                &LPIPS&0.0001&0.0002&0.0003&0.0322&0.0036&-\\
        \bottomrule
    \end{tabular}}
\end{table*}
%----------------------------------------------------------------------------------
\subsubsection{Frechet Inception Distance}
The FID performance of fine-tuned model on the research of Chen \etal \cite{28} is treated as the baseline for evaluation. In addition, we conducted training for the DDPM model while assessing the FID of images generated from noise, following a methodology akin to their prior research. The corresponding FID results for our experiments, encompassing both the protected dataset $\mathcal{D}_1^p$ and the recovered dataset $\mathcal{D}_1^{s n}$ at the proposed permission level 1, have been summarized in Table \ref{tab:fidchenandours}.
% In the case of Clean $\mathcal{D}_1^{s n}$, FID is as low as 0.33 and 0.69 on CIFAR-10 and CelebA, respectively (with an optimal FID of 0). In contrast, protected $\mathcal{D}_1^p$ has significantly higher FID of 85.48 and 62.63 on CIFAR-10 and CelebA. After the restoration, recovered $\Tilde{\mathcal{D}}_1^{s n}$ achieved notably lower FID scores, measuring 1.00 and 3.92 respectively.
% Additionally, we train the DDPM with the FID of generated images from noise similar to their work for our experiments whose FID are shown in Table \hyperref[tab:fidchenandours]{3}. The FID for the protected dataset $\mathcal{D}_1^p$ and the recovered dataset $\mathcal{D}_1^{s n}$ at the proposed permission level 1 are presented in Table \hyperref[tab:fidlevel1]{4}. 
The observation of a significant impact on the FID of the protected dataset $\mathcal{D}_1^p$ indicates that AIGC models are difficult to train on these unrealistic datasets. Nevertheless, despite this influence, the recovered dataset $\Tilde{\mathcal{D}}_1^{s n}$ retains the original performance to a great extent.
%\definecolor{mygray}{RGB}{211,210,208}
\begin{table}[htbp]
    \centering
    \renewcommand{\arraystretch}{1}
    \caption{The FID of generated images from noise along with $\mathcal{D}_1^{s n}$, $\mathcal{D}_1^p$ and $\Tilde{\mathcal{D}}_1^{s n}$ on CIFAR-10 and CelebA.}
    \label{tab:fidchenandours}
    \resizebox{0.6\textwidth}{!}{
    \begin{tabular}{c|c|cc}
         \toprule
         \multicolumn{2}{c|}{\multirow{1}{*}{\textbf{Method}}}
         & {{\textbf{CIFAR-10}}} & {{\textbf{CelebA}}} \\
         \midrule
         Chen \etal \cite{28}& From noise &4.74  & 5.44 \\
         \midrule
         \multirow{4}{*}{Ours}
         &From noise& 4.59 & 5.60\\
         
         &Clean $\mathcal{D}_1^{s n}$& 0.33 & 0.69\\
         
         &Protected $\mathcal{D}_1^p$& 85.48 & 62.63\\
         
         &Recovered $\Tilde{\mathcal{D}}_1^{s n}$& 1.00 & 3.92\\
         \bottomrule
    \end{tabular}}
    
\end{table}
%----------------------------------------------------------------------------------
\subsubsection{Robustness}
\begin{figure}[h]
    \centering
    \includegraphics[width=0.5\linewidth]{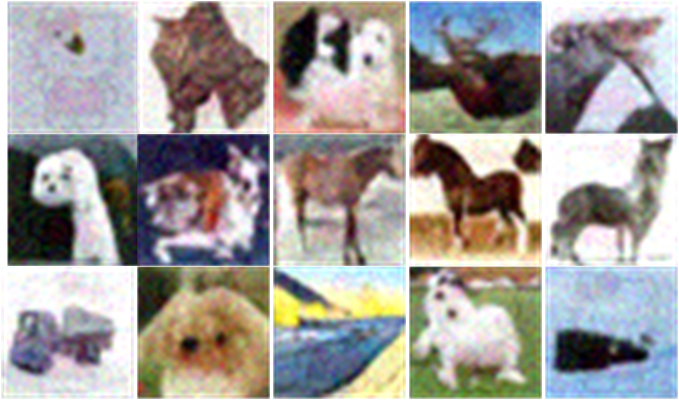}
    \caption{Visualization of images generated from attacker's DDPM. The DDPM trained on the protected dataset has met a significant performance degradation. Additionally, it is able to discern the trigger in the generated images.}
    \label{fig:attacker fid}
\end{figure}

\textbf{RAEDiff} effectively maintains its confidentiality. However, we also need to consider the possibility of attacks, where attackers have a comprehensive understanding of our protection mechanisms attempting to restore the images. To address this concern, we conducted a simulation experiment. We assumed that attackers had managed to steal the protected dataset $\mathcal{D}_1^p$ from authorized users. Moreover, they possess hardware and software conditions similar to ours. With these resources, attackers trained an equivalent DDPM using $\mathcal{D}_1^p$. The only difference is that their model does not introduce any bias to the Gaussian distribution but instead uses the original standard Gaussian distribution. After illegal training, they get $\Tilde{\mathcal{D}}_1^{a t k}$, the performance of $\Tilde{\mathcal{D}}_1^{a t k}$ is shown in the last column of Table \ref{tab:ssimpsnrlpips}. It can be observed that due to the profound impact of the protected dataset on the AIGC model, even if an attacker were to train a DDPM using the same approach as ours, they would be unable to restore the original images. Instead, this would lead to further degradation in image quality.
Meanwhile, the generated images from noises on attackers' DDPM are shown in Figure \ref{fig:attacker fid}, demonstrating that our protected dataset similarly exerts a significant influence on the AIGC model.
Hence, even when equipped with knowledge about our method, attackers are unable to effectively recover the images.%Meanwhile, the generated images from noises on attackers' DDPM are shown in Figure  \hyperref[fig:attacker fid]{4}, demonstrating that our protected dataset similarly exerts a significant influence on the AIGC model.
% \begin{table}[htbp]
%     \centering
%     \renewcommand{\arraystretch}{1}
%     \caption{The Performance of $\mathcal{D}_1^p$ and $\Tilde{\mathcal{D}}_1^{a t k}$ on CIFAR-10.}
%     \label{tab:attacker}
%     \resizebox{0.32\textwidth}{!}{
%     \begin{tabular}{c|cc}
%          \toprule
%           \textbf{Metrics}&{\textbf{$\mathcal{D}_1^p$}} & {{\textbf{$\Tilde{\mathcal{D}}_1^{a t k}$}}} \\
%          \midrule
%           SSIM & 0.832 & 0.787\\
%           PSNR & 31.79 & 27.96\\
%           LPIPS& 0.0086 & 0.0121\\
%          \bottomrule
%     \end{tabular}
%     }

\subsubsection{Theoretical reversibility}
\begin{figure*}[htbp]
\centering
\begin{tabular}{ccc}
  \toprule
  $\mathcal{D}$ & $\Tilde{\mathcal{D}}_1^{s n}$ &  $\mathcal{D}_1^p$\\
  \midrule
  \includegraphics[width=0.32\linewidth]{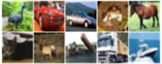}
  &
  \includegraphics[width=0.32\linewidth]{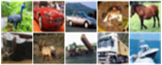}  
&
  \includegraphics[width=0.32\linewidth]{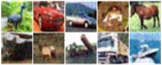}\\
  \bottomrule
\end{tabular}
\label{fig:wanmeidetu}
\caption{Visualization of $\mathcal{D}$, $\Tilde{\mathcal{D}}_1^{s n}$ and $\mathcal{D}_1^p$ on CIFAR-10 under ideal conditions. The SSIM, PSNR, and LPIPS values for $\Tilde{\mathcal{D}}_1^{s n}$ are 0.995, 36.68, and 0.0003, respectively.}
\end{figure*}

We exclusively trained the DDPM on a subset of 50 CIFAR-10 images to explore the theoretical reversibility achievable under ideal conditions using our proposed method.
As shown in %Table \hyperref[tab:wanmei]{8} and 
Figure \hyperref[fig:wanmeidetu]{8}, \textbf{RAEDiff} could be used solely through the DDPM's prior knowledge of the data to recover $\mathcal{D}_1^p$ to $\Tilde{\mathcal{D}}_1^{s n}$,  achieving great self-recovery after fully fitting the data.
%The SSIM, PSNR, and LPIPS values for $\Tilde{\mathcal{D}}_1^{s n}$ are 0.995, 36.68, and 0.0003, respectively.
% \begin{figure}[htbp]
% \centering
% \begin{tabular}{c}
%   \toprule
%   $\mathcal{D}$ \\
%   \midrule
%   \includegraphics[width=0.9\linewidth]{wanmeiyuantu.png} \\  
%   \toprule
%   $\Tilde{\mathcal{D}}_1^{s n}$\\
%   \midrule
%   \includegraphics[width=0.9\linewidth]{wanmeidsnbolang.png} \\
%   \toprule
%    $\mathcal{D}_1^p$\\
% \midrule
% \includegraphics[width=0.9\linewidth]{wanmeidp.png}\\
%   \bottomrule
% \end{tabular}
% \label{fig:wanmeidetu}
% \caption{Visualization of $\mathcal{D}$, $\Tilde{\mathcal{D}}_1^{s n}$ and $\mathcal{D}_1^p$ on CIFAR-10 under ideal conditions. The SSIM, PSNR, and LPIPS values for $\Tilde{\mathcal{D}}_1^{s n}$ are 0.995, 36.68, and 0.0003, respectively.}
% \end{figure}
\section{Conclusion}
\label{sec:conclu}
In this paper, a self-generation and self-recovery RAE method based on DDPM with a biased Gaussian distribution is proposed.
Experiments are conducted to examine the impact of RAEs on the training performance of the AIGC model for the first time.
Compared to existing RAE methods, our proposed method has a significant impact on model testing accuracy, resulting in a substantial increase (e.g., 42.75\% testing accuracy on CIFAR-10). Furthermore, it first effectively restores RAEs to the original image without any auxiliary information(e.g., SSIM of 0.969 on CIFAR-10). The model accuracy and visual evaluation metrics results demonstrate that our proposed method effectively introduces adversarial perturbations to both the classification model and the AIGC model.
In future research, we intend to optimize the AGP in the diffusion model and explore the possibility of achieving constraints through loss functions to enhance the adversarial attack strength of RAEs. Furthermore, we aim to examine the impact of various network structures on the diffusion model to enhance its ability to refine the introduced noise on RAEs for superior visual quality.

% 明天增加迁移性实验，以及和他们的对比(还没加
% 明天加motivated by diffusion后门攻击，以及diffusion模型版权保护的研究和diffusion model的一些研究，发现扩散模型在补充图像语义信息方面有优秀得效果(deal)
% \section*{Acknowledgments}
% This work was supported by Hainan Province Key R\&D plan project (No.ZDYF2022GXJS224)
%\clearpage
\bibliographystyle{splncs04}
\bibliography{main}

\begin{thebibliography}{10}
\providecommand{\url}[1]{\texttt{#1}}
\providecommand{\urlprefix}{URL }
\providecommand{\doi}[1]{https://doi.org/#1}

\bibitem{2}
Banesh, D., Petersen, M.R., Ahrens, J., Turton, T.L., Samsel, F., Schoonover, J., Hamann, B.: An image-based framework for ocean feature detection and analysis. Journal of Geovisualization and Spatial Analysis  \textbf{5},  1--21 (2021)

\bibitem{32}
Chen, C., Wu, Z., Lai, Y., Ou, W., Liao, T., Zheng, Z.: Challenges and remedies to privacy and security in aigc: Exploring the potential of privacy computing, blockchain, and beyond. arXiv preprint arXiv:2306.00419  (2023)

\bibitem{28}
Chen, W., Song, D., Li, B.: Trojdiff: Trojan attacks on diffusion models with diverse targets. In: Proceedings of the IEEE/CVF Conference on Computer Vision and Pattern Recognition. pp. 4035--4044 (2023)

\bibitem{29}
Chou, S.Y., Chen, P.Y., Ho, T.Y.: How to backdoor diffusion models? In: Proceedings of the IEEE/CVF Conference on Computer Vision and Pattern Recognition. pp. 4015--4024 (2023)

\bibitem{9}
Dai, L., Mao, J., Xu, L., Fan, X., Zhou, X.: Balancing robustness and covertness in nlp model watermarking: A multi-task learning approach. In: 2023 IEEE Symposium on Computers and Communications (ISCC). pp. 1376--1382. IEEE (2023)

\bibitem{6}
Fan, X., Zhou, X., Zhu, B., Dong, J., Niu, J., Wang, H.: Survey of copyright protection schemes based on dnn model. Journal of Computer Research and Development  \textbf{59}(2022-05-953),  953--977 (2022)

\bibitem{13}
Fu, D., Zhou, X., Xu, L., Hou, K., Chen, X.: Robust reversible watermarking by fractional order zernike moments and pseudo-zernike moments. IEEE Transactions on Circuits and Systems for Video Technology  (2023)

\bibitem{50res18}
He, K., Zhang, X., Ren, S., Sun, J.: Deep residual learning for image recognition. In: Proceedings of the IEEE conference on computer vision and pattern recognition. pp. 770--778 (2016)

\bibitem{41}
Heusel, M., Ramsauer, H., Unterthiner, T., Nessler, B., Hochreiter, S.: Gans trained by a two time-scale update rule converge to a local nash equilibrium. Advances in neural information processing systems  \textbf{30} (2017)

\bibitem{40}
Ho, J., Jain, A., Abbeel, P.: Denoising diffusion probabilistic models. Advances in neural information processing systems  \textbf{33},  6840--6851 (2020)

\bibitem{38}
Krizhevsky, A., Hinton, G., et~al.: Learning multiple layers of features from tiny images  (2009)

\bibitem{45}
Kumar, S., Gupta, A., Walia, G.S.: Reversible data hiding: A contemporary survey of state-of-the-art, opportunities and challenges. Applied Intelligence pp. 1--34 (2022)

\bibitem{1}
Li, P., Tu, S., Xu, L.: Deep rival penalized competitive learning for low-resolution face recognition. Neural Networks  \textbf{148},  183--193 (2022)

\bibitem{36}
Liu, J., Zhang, W., Fukuchi, K., Akimoto, Y., Sakuma, J.: Unauthorized ai cannot recognize me: Reversible adversarial example. Pattern Recognition  \textbf{134},  109048 (2023)

\bibitem{39}
Liu, Z., Luo, P., Wang, X., Tang, X.: Deep learning face attributes in the wild. In: Proceedings of the IEEE international conference on computer vision. pp. 3730--3738 (2015)

\bibitem{7}
Long, D., Jing, Z., Xuefeng, F., Xiaoyi, Z.: Nlp neural network copyright protection based on black box watermark. Chinese Journal of Network \& Information Security  \textbf{9}(1) (2023)

\bibitem{52repaint}
Lugmayr, A., Danelljan, M., Romero, A., Yu, F., Timofte, R., Van~Gool, L.: Repaint: Inpainting using denoising diffusion probabilistic models. In: Proceedings of the IEEE/CVF Conference on Computer Vision and Pattern Recognition. pp. 11461--11471 (2022)

\bibitem{18}
Mao, C., Chiquier, M., Wang, H., Yang, J., Vondrick, C.: Adversarial attacks are reversible with natural supervision. In: Proceedings of the IEEE/CVF International Conference on Computer Vision. pp. 661--671 (2021)

\bibitem{53ddpm}
Nair, N.G., Mei, K., Patel, V.M.: At-ddpm: Restoring faces degraded by atmospheric turbulence using denoising diffusion probabilistic models. In: Proceedings of the IEEE/CVF Winter Conference on Applications of Computer Vision. pp. 3434--3443 (2023)

\bibitem{47}
Nguyen, A., Tran, A.: Wanet--imperceptible warping-based backdoor attack. arXiv preprint arXiv:2102.10369  (2021)

\bibitem{3}
Ning, X., Tian, W., Yu, Z., Li, W., Bai, X., Wang, Y.: Hcfnn: high-order coverage function neural network for image classification. Pattern Recognition  \textbf{131},  108873 (2022)

\bibitem{49}
Ribeiro, M., Grolinger, K., Capretz, M.A.: Mlaas: Machine learning as a service. In: 2015 IEEE 14th international conference on machine learning and applications (ICMLA). pp. 896--902. IEEE (2015)

\bibitem{48}
Salem, A., Wen, R., Backes, M., Ma, S., Zhang, Y.: Dynamic backdoor attacks against machine learning models. In: 2022 IEEE 7th European Symposium on Security and Privacy (EuroS\&P). pp. 703--718. IEEE (2022)

\bibitem{55ssim}
Wang, Z., Bovik, A.C., Sheikh, H.R., Simoncelli, E.P.: Image quality assessment: from error visibility to structural similarity. IEEE transactions on image processing  \textbf{13}(4),  600--612 (2004)

\bibitem{5}
Wei, T., Chen, D., Zhou, W., Liao, J., Tan, Z., Yuan, L., Zhang, W., Yu, N.: Hairclip: Design your hair by text and reference image. In: Proceedings of the IEEE/CVF Conference on Computer Vision and Pattern Recognition. pp. 18072--18081 (2022)

\bibitem{22}
Xiong, L., Wu, Y., Yu, P., Zheng, Y.: A black-box reversible adversarial example for authorizable recognition to shared images. Pattern Recognition  \textbf{140},  109549 (2023)

\bibitem{21}
Xue, M., Wu, Y., Zhang, Y., Wang, J., Liu, W.: Dataset authorization control: protect the intellectual property of dataset via reversible feature space adversarial examples. Applied Intelligence  \textbf{53}(6),  7298--7309 (2023)

\bibitem{4}
Yang, J., Li, C., Zhang, P., Xiao, B., Liu, C., Yuan, L., Gao, J.: Unified contrastive learning in image-text-label space. In: Proceedings of the IEEE/CVF Conference on Computer Vision and Pattern Recognition. pp. 19163--19173 (2022)

\bibitem{20}
Yin, Z., Chen, L., Lyu, W., Luo, B.: Reversible attack based on adversarial perturbation and reversible data hiding in yuv colorspace. Pattern Recognition Letters  \textbf{166}, ~1--7 (2023)

\bibitem{19}
Yin, Z., Wang, H., Chen, L., Wang, J., Zhang, W.: Reversible adversarial attack based on reversible image transformation. arXiv preprint arXiv:1911.02360  (2019)

\bibitem{44}
Zhang, R., Isola, P., Efros, A.A., Shechtman, E., Wang, O.: The unreasonable effectiveness of deep features as a perceptual metric. In: Proceedings of the IEEE conference on computer vision and pattern recognition. pp. 586--595 (2018)

\bibitem{51rit}
Zhang, W., Wang, H., Hou, D., Yu, N.: Reversible data hiding in encrypted images by reversible image transformation. IEEE Transactions on multimedia  \textbf{18}(8),  1469--1479 (2016)

\bibitem{14}
Zhou, X., Hou, K., Zhuang, Y., Yin, Z., Han, W.: General pairwise modification framework for reversible data hiding in jpeg images. IEEE Transactions on Circuits and Systems for Video Technology  (2023)

\end{thebibliography}

%\input{sec/X_biography}
\end{document}